\documentclass[amssymb, amsmath,nofootinbib, aps, pra, superscriptaddress]{revtex4-2}
\usepackage{graphicx}
\usepackage{verbatim}
\usepackage{xcolor}
\usepackage{esint}
\usepackage{dsfont}
\usepackage[T1]{fontenc}
\usepackage[utf8]{inputenc}
\usepackage[hidelinks]{hyperref}
\begin{document}
\title{Local Scale Invariance in Quantum Theory: Experimental Predictions}
\author{Indrajit Sen}
\email{isen@ggc.edu}
\affiliation{Department of Physics and Pre-Engineering, Georgia Gwinnett College\\
1000 University Center Lane, Lawrenceville, GA, 30043, USA}
\affiliation{Institute for Quantum Studies, Chapman University\\
One University Drive, Orange, CA, 92866, USA}
\author{Matthew Leifer}
\affiliation{Institute for Quantum Studies, Chapman University\\
One University Drive, Orange, CA, 92866, USA}
\affiliation{Schmid College of Science and Technology, Chapman University\\
One University Drive, Orange, CA, 92866, USA}

\begin{abstract}
We explore the experimental predictions of the local scale invariant, non-Hermitian pilot-wave (de Broglie-Bohm) formulation of quantum theory introduced in arXiv:2601.03567. We use Weyl's definition of gravitational radius of charge to obtain the fine-structure constant for non-integrable scale effects $\alpha_S$. The minuteness of $\alpha_S$ relative to $\alpha$ ($\alpha_S/\alpha \sim 10^{-21}$) effectively hides the effects in usual quantum experiments. In an Aharonov-Bohm double-slit experiment, the theory predicts that the position probability density depends on which slit the particle trajectory crosses, due to a non-integrable scale induced by the magnetic flux. This experimental prediction can be tested for an electrically neutral, heavy molecule with mass $m \sim 10^{-19} \text{g}$ at a $\sim 10^5 \text{ esu}$ flux regime. We analyse the Weyl-Einstein debate on the second-clock effect using the theory and show that spectral frequencies are history-independent. We thereby resolve Einstein's key objection against local scale invariance, and obtain two further experimental predictions. First, spectral intensities turn out to be history-dependent. Second, energy eigenvalues are modified by tiny imaginary corrections that modify spectral linewidths. We argue that the trajectory dependence of the probabilities renders our theory empirically distinguishable from other quantum formulations that do not use pilot-wave trajectories, or their mathematical equivalents, to derive experimental predictions. 
\end{abstract}

\maketitle

\section{Introduction}
It is well understood that quantum states are rays, that is, they can be multiplied by an arbitrary complex constant without affecting any physical predictions. However, it is not clearly understood why the phase and the magnitude of the constant exhibit markedly different behaviors when one attempts to localize them. The phase of the constant can be promoted to a local parameter, giving rise to local phase invariance with gauge group $U(1)$ -- one of the fundamental gauge symmetries in the Standard Model \cite{macha, pesky}. In contrast, the magnitude of the constant cannot be promoted to a local degree of freedom without breaking the unitary structure of orthodox quantum theory. Unlike local phase invariance, local scale invariance symmetry with gauge group $\mathbb{R}^+$ is considered fundamentally incompatible with quantum theory \cite{gawn}. \\

In a previous paper \cite{sen26a}, we showed that this disparity in the status of these symmetries is an artifact of orthodox quantum theory that vanishes in the pilot-wave formulation. We showed that the structure of pilot-wave theory\footnote{Also known as de Broglie-Bohm theory, Bohmian mechanics or the Causal interpretation in the literature.} (PWT) naturally admits a complex electromagnetic gauge coupling parameter $e_C = e + ie_I $, whose imaginary component $e_I$ implements local scale invariance. The resultant theory is non-Hermitian, with the conserved current density given by the local scale invariant, trajectory-dependent ratio $|\psi|^2/\mathds 1^2[\mathcal C]$, where $\mathds 1[\mathcal C]$ is a scale factor along the pilot-wave trajectory $\mathcal C$ in configuration space. The Born rule density $|\psi|^2$ is recovered in the special case when non-integrable scale changes can be ignored, so that $\mathds 1[\mathcal C] $ can be set equal to 1 everywhere. \\

In this article, we explore the experimental predictions of non-Hermitian PWT to, first, test the theory, and second, distinguish it from other quantum formulations. We first estimate the value of $e_I$, which determines the magnitude of non-integrable scale effects, in section \ref{value}. We then discuss a key testable experimental prediction of our theory in an Aharonov-Bohm setup in section \ref{expt}. We analyse the Weyl-Einstein debate on the second-clock effect \cite{vizgin, ryckmanbook} using our theory in section \ref{wade}. We show that Einstein's criticism of history-dependent spectral frequencies is incorrect in \ref{specter}. Our analysis yields two further theoretical predictions that we discuss in sections \ref{intense} and \ref{physical}. First, spectral intensities are history-dependent. Second, spectral linewidths are modified by the contributions of imaginary components of energy eigenvalues. We discuss our results in section \ref{disc}.

\section{Estimate of the imaginary component of the gauge coupling parameter} \label{value}
Weyl's unified field theory \cite{weyl18, weyl18a, weyl19, weyl22} predicts that a vector of initial magnitude $l_0$ transported around a closed loop $\mathcal C$, upon returning to its initial position, has a length 
\begin{align}
l \equiv l_0 e^{-\frac{1}{\gamma}\oint_{\mathcal C} A^\mu dx_\mu} \label{len}
\end{align}
where $A^\mu$ is the electromagnetic gauge field and $\gamma$ is a constant that has the same dimension as $\oint_{\mathcal C} A^\mu dx_\mu$, which is charge or flux in CGS units\footnote{Charge and flux have the same dimensions in CGS units.}. To obtain clear experimental predictions, we first need to determine the value of $\gamma$. A definite value for $\gamma$ may appear philosophically at odds with the spirit of Weyl's theory as it sets a fixed scale for the magnitude of non-integrable scale effects. Weyl argued that although there are no absolute scales in the theory, certain combinations of physical quantities can define relational, effective length scales \cite{weyl19, weyl22}. These length scales then allow the definition of practical units of measurement in a fundamentally scale-invariant theory. This is essentially the approach that we take to estimate the value of $\gamma$, and thereby the value of the imaginary gauge coupling parameter $e_I$, below.\\

We start with the adaption of Weyl's ideas to orthodox quantum theory, whereupon the change in length of the vector (amplitude of the wavefunction) is replaced by the change in the direction of the vector (phase of the wavefunction) \cite{fuck, london27}. The phase factor gained by the wavefunction of an electron around a loop $\mathcal C$ is given by
\begin{align}
    e^{i \alpha\frac{\oint_\mathcal{C}  A^\mu dx_\mu}{e}} \label{fine}
\end{align}
where $\alpha$ is the fine-structure constant and $e$ is the electronic charge. The exponent in (\ref{fine}) can be physically interpreted as the magnitude of the flux $\oint_\mathcal{C}  A^\mu dx_\mu$, measured with respect to the fundamental unit of charge $e$, multiplied by the fine-structure constant $\alpha$, which is a dimensionless constant that determines the magnitude of the phase change for a given magnitude of flux. The fine-structure constant can be defined as the ratio of two characteristic lengths scales:
\begin{align}
\alpha \equiv \frac{r_e}{\lambdabar} \label{alpha}
\end{align}
where $r_e \equiv e^2/m_e c^2$ is the electron's Lorentz radius (or Thomson scattering radius), $\lambdabar \equiv \hbar/m_e c$ is the electron's reduced Compton wavelength and $m_e$ is the electron's mass. Thus, $\alpha$ can be physically interpreted as the measure of the electron's characteristic electromagnetic length scale with respect to its quantum mechanical length scale.\\

The above suggests that we may analogously posit the constant $\gamma$ in (\ref{len}) to be 
\begin{align}
    \gamma \equiv \frac{e}{\alpha_S} \label{vasco}
\end{align}
where $e$ is the fundamental unit of charge (electronic charge) and $\alpha_S$ is the analogue of $\alpha$ for non-integrable scale effects. That is, $\alpha_S$ is a dimensionless constant that determines the magnitude of the change in amplitude of the wavefunction for a given amount of flux. Analogous to $\alpha$, we define $\alpha_S$ as the ratio of the electron's gravitational radius due to its charge and its Compton wavelength. The gravitational radius of a charge is the length scale at which the charge's scale effects on the spacetime manifold dominate \cite{weyl19,weyl22}. It is defined as
\begin{align}
    r_g \equiv \frac{e\sqrt G}{c^2} \label{radius}
\end{align}
for charge $e$, where $G$ is Newton's gravitational constant. This implies that
\begin{align}
    \alpha_S = \frac{r_g}{\lambdabar} = \frac{em_e\sqrt G}{c\hbar} \label{finescale}
\end{align}
Note that $\alpha_S$ can be expressed as
\begin{align}
    \alpha_S = \sqrt{\alpha \times \alpha_G} \label{expres}
\end{align}
where $\alpha_G \equiv Gm_e^2/\hbar c$ is the gravitational fine-structure constant, defined as the ratio of the gravitational radius of the electron's mass $Gm_e/c^2$ \cite{weyl22} and its Compton wavelength. The expression \eqref{expres} makes sense as the gravitational and electromagnetic fields are unified in Weyl's theory as part of the spacetime manifold. The constant $\gamma$ in (\ref{len}) can be found from equations (\ref{vasco}), (\ref{finescale}): 
\begin{align}
    \gamma = \frac{c\hbar}{m_e \sqrt G} \label{da}
\end{align}
Equation (\ref{da}) implies that the change in scale of the wavefunction is independent of the electronic charge $e$, analogous to the change in phase of the wavefunction being independent of the electronic mass $m_e$. Further, it implies that the change in length is a gravito-electromagnetic effect where the electron's mass $m_e$ couples to the electromagnetic field $A^\mu$. Thus, the electron couples to the electromagnetic gauge field in two distinct ways in our theory: dynamically via its charge, and geometrically via its mass. This is surprising but makes sense in the context of Weyl's unified field theory, where the electromagnetic gauge field is part of the spacetime manifold and affects all particles regardless of their charge by affecting the spacetime scaling.\\

Lastly, we know that for a particle with arbitrary charge $q$, the change in phase of the wavefunction is given by $\alpha(q/e) (\oint_\mathcal{C}  A^\mu dx_\mu/e)$. This suggests by analogy that, for a particle with arbitrary mass $m$, the change in scale is given by $\alpha_S(m/m_e) (\oint_\mathcal{C}  A^\mu dx_\mu/e)$. As the scale factor can be written in terms of the imaginary component of the gauge coupling parameter $e_I$ as $e^{-(e_I/\hbar c)\oint_\mathcal{C}  A^\mu dx_\mu}$, we find 
\begin{align}
    e_I \equiv m \sqrt{G} \label{mast}
\end{align}
for an arbitrary quantum particle of mass $m$. \\

Using the numerical values of the constants $e, G, c$ in (\ref{radius}), we get 
\begin{align}
    \frac{\alpha_S}{\alpha} = \frac{m_e\sqrt G}{e} \approx 4.9 \times 10^{-22} 
\end{align}
Therefore, the change in the scale of the wavefunction is expected to be $\sim 10^{-22}$ times smaller than the change in the phase of the wavefunction. This helps explain why non-integrable scale effects have not been observed in quantum experiments till date, if non-Hermitian PWT is realized in nature. The value of the imaginary gauge coupling parameter for the electron is (in CGS units)
\begin{align}
    e_I = m_e \sqrt{G} \approx 2.35 \times 10^{-31} \text{esu}
\end{align}
which is $\sim 10^{-21}$ times smaller than the electronic charge $e = 4.8 \times 10^{-10} \text{esu}$.  

\section{Testable prediction in an Aharonov-Bohm setup} \label{wab}
We propose an experiment in this section that can test non-Hermitian PWT and distinguish it from other quantum formulations. We start by discussing the effect of $e_I$ on the equilibrium position probability density in an Aharonov-Bohm (AB) experiment \cite{ahoaho, abreview, peshki} (also known as Ehrenberg–Siday–Aharonov–Bohm experiment \cite{thahro, ahoaho2}).
\subsection{Which-way-dependent probabilities in Aharonov-Bohm double-slit experiment} \label{expt}
The introduction of $e_I$ leads to a small, but in principle observable, change in the AB effect. We present our argument using the vector potential form of the AB effect in a double-slit experiment, but it can be straightforwardly generalized. In the experiment, the wavepacket of a charged quantum particle is split by the two slits into partial wavepackets that propagate on opposite sides of a solenoid, placed immediately behind the slits, enclosing magnetic flux. The partial wavepackets propagate in flux-free regions and coherently recombine to yield an interference pattern that is shifted due to the flux. For a discussion of this effect from a pilot-wave perspective, see \cite{kaye, bohmbook2, hollandbook}.\\

Suppose the charged particle is an electron and the two slits are labelled A and B. Let the partial wavepackets emerging from A, B at the time of their recombination, with the solenoid switched off, be described by wavefunctions $\psi_A(\vec x, t)$, $\psi_B(\vec x, t)$ respectively. Then, for the solenoid switched on, the corresponding partial wavefunctions are given by
\begin{align}
    \psi'_A(\vec x, t) = \psi_A(\vec x, t) e^{i\frac{e}{\hbar c}\int^{\vec{x}}_{\mathcal{C}_A} \vec A(\vec x') \cdot d \vec x' }e^{-\frac{e_I}{\hbar c}\int^{\vec{x}}_{\mathcal{C}_A} \vec A(\vec x') \cdot d \vec x' } \label{switch1}\\
    \psi'_B(\vec x, t) = \psi_B(\vec x, t) e^{i\frac{e}{\hbar c}\int^{\vec{x}}_{\mathcal{C}_B} \vec A(\vec x') \cdot d \vec x'} e^{-\frac{e_I}{\hbar c}\int^{\vec{x}}_{\mathcal{C}_B} \vec A(\vec x') \cdot d \vec x' } \label{switch2}
\end{align}
where $\mathcal{C}_A$ ($\mathcal{C}_B$) labels a path, lying outside the solenoid, from the preparation source to $(\vec x, t)$ through slit $A$ ($B$). Clearly, the partial wavepackets not only accumulate different phases along their paths, but also different amplitudes, as $e_C = e + ie_I$ is complex. \\

Suppose, when the solenoid is switched off, the recombined wavefunction is given by $\big (\psi_A(\vec x, t) + \psi_B(\vec x, t)\big)/\sqrt{2}$. Then, with the solenoid switched on, the recombined wavefunction is given by $\psi'_{AB}(\vec x, t) = \big (\psi'_A(\vec x, t) + \psi'_B(\vec x, t)\big )/\sqrt{2}$, which can be written in two different ways:
\begin{align}
    \psi'_{AB}(\vec x, t) &=\frac{\big (\psi_A(\vec x, t) + \psi_B(\vec x, t)e^{i\frac{e}{\hbar c}\oint_{\mathcal{C}_L} \vec A(\vec x') \cdot d \vec x'}e^{-\frac{e_I}{\hbar c}\oint_{\mathcal{C}_L} \vec A(\vec x') \cdot d \vec x'}\big )}{\sqrt{2}} e^{i\frac{e}{\hbar c}\int^{\vec{x}}_{\mathcal{C}_A} \vec A(\vec x') \cdot d \vec x' }e^{-\frac{e_I}{\hbar c}\int^{\vec{x}}_{\mathcal{C}_A} \vec A(\vec x') \cdot d \vec x' } \nonumber\\
    &= \frac{\big (\psi_A(\vec x, t)e^{-i\frac{e}{\hbar c}\oint_{\mathcal{C}_L} \vec A(\vec x') \cdot d \vec x'}e^{\frac{e_I}{\hbar c}\oint_{\mathcal{C}_L} \vec A(\vec x') \cdot d \vec x'} + \psi_B(\vec x, t)\big )}{\sqrt{2}} e^{i\frac{e}{\hbar c}\int^{\vec{x}}_{\mathcal{C}_B} \vec A(\vec x') \cdot d \vec x' }e^{-\frac{e_I}{\hbar c}\int^{\vec{x}}_{\mathcal{C}_B} \vec A(\vec x') \cdot d \vec x' } \label{ab}
\end{align}
where we have used equations (\ref{switch1}), (\ref{switch2}) and $\mathcal{C}_L$ is a closed loop around the solenoid such that $\int^{\vec{x}}_{\mathcal{C}_B} \vec A(\vec x') \cdot d \vec x'  - \int^{\vec{x}}_{\mathcal{C}_A} \vec A(\vec x') \cdot d \vec x'  = \oint_{\mathcal{C}_L} \vec A(\vec x') \cdot d \vec x'$. \\
\begin{figure}
    \centering
    \includegraphics[scale=0.7]{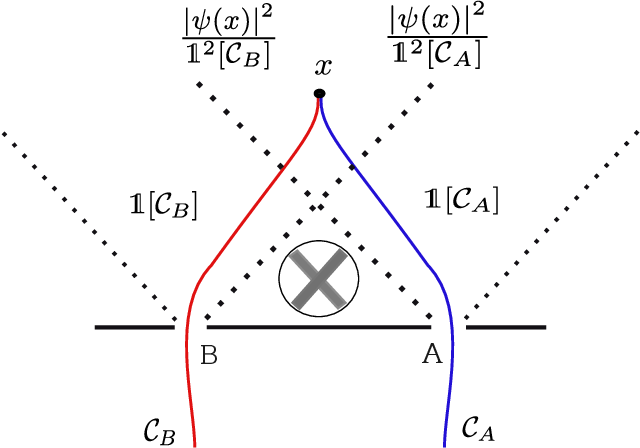}
    \caption{Schematic illustration of the which-way-dependent probability density in a double-slit Aharonov-Bohm experiment. Two partial wavepackets (shown by dotted lines) emerging from slits A and B pass on either side of a solenoid enclosing magnetic flux (shown by $\otimes$) and then coherently recombine. A spacetime point $x \equiv (ct, \vec x)$ in the support of the recombined wavepacket is indicated. Two schematic paths $\mathcal C_A$ and $\mathcal C_B$ (lying outside the solenoid) connecting the preparation source to $x$ via slit A and B are indicated in \textcolor{blue}{blue} and \textcolor{red}{red}, respectively. If the particle trajectory passing through $x$, as determined by the pilot-wave guidance equation, crosses slit A (B), then non-Hermitian PWT predicts the equilibrium probability density at $x$ to be $|\psi(x)|^2/\mathds 1^2[\mathcal C_A]$ $\big (|\psi( x)|^2/\mathds 1^2[\mathcal C_B]\big )$. Note that the guidance equation associates any spacetime point in the support of the recombined wavepacket with a single particle trajectory crossing either slit A or B.} 
    \label{abf}
\end{figure}

The equilibrium probability density is given by $|\psi'_{AB}(\vec x, t)|^2/\mathds 1^2[\mathcal C]$, where $\mathds 1[\mathcal C]= e^{-\frac{e_I}{\hbar c} \int^{\vec{x}}_{\mathcal{C}} \vec A(\vec x') \cdot d \vec x' }$ is the scale factor evaluated along the particle trajectory $\mathcal C$ passing through $(\vec x, t)$ \cite{sen26a}. As the partial wavepackets propagate in flux-free regions, the scale factor $\mathds 1[\mathcal C]$ does not depend on the details of the trajectories\footnote{From Stokes' theorem, the scale factor $\mathds 1[\mathcal C]= e^{-\frac{e_I}{\hbar c} \int^{\vec{x}}_{\mathcal{C}} \vec A(\vec x') \cdot d \vec x' }$ depends on only the end points of the trajectory $\mathcal C$, if $\mathcal C$ lies within a region where $\vec \nabla \times \vec A = 0$.}. Therefore, if the particle has travelled via slit A (B) to reach the spacetime point $(\vec x, t)$, then $\mathds 1[\mathcal C] = \mathds 1[\mathcal C_A]$ ($\mathds 1[\mathcal C] = \mathds 1[\mathcal C_B]$). This implies that the equilibrium probability density at $(\vec x, t)$ is
\begin{align}
   \frac{|\psi'_{AB}(\vec x, t)|^2}{\mathds 1^2[\mathcal C]} = \left \{ \begin{array}{cc}
        |\psi_A(\vec x, t) + \psi_B(\vec x, t)e^{i\frac{e}{\hbar c}\oint_{\mathcal{C}_L} \vec A(\vec x') \cdot d \vec x'}e^{-\frac{e_I}{\hbar c}\oint_{\mathcal{C}_L} \vec A(\vec x') \cdot d \vec x'}|^2/2&,\text{ if the particle trajectory crosses slit A} \\
        \textbf{ } & \textbf{ } \\
        |\psi_A(\vec x, t)e^{-i\frac{e}{\hbar c}\oint_{\mathcal{C}_L} \vec A(\vec x') \cdot d \vec x'}e^{\frac{e_I}{\hbar c}\oint_{\mathcal{C}_L} \vec A(\vec x') \cdot d \vec x'} + \psi_B(\vec x, t)|^2/2  &,\text{ if the particle trajectory crosses slit B}    \end{array} \right . \label{wow}
\end{align}
where we have used (\ref{ab}). Let us make a few observations on the probability density given by equation (\ref{wow}). We first note that it is gauge invariant. Second, the probability density at $(\vec x, t)$ explicitly depends on which slit the particle trajectory\footnote{Note that approximate trajectories followed by narrowly peaked wavepackets, or trajectories in Feynman's path-integral formulation, are irrelevant to the discussion here. For an introduction to pilot-wave trajectories, see \cite{bohmbook2, hollandbook}.} passing through $(\vec x, t)$ crosses. This can only make sense in a theory with definite particle trajectories in the first place. Further, the trajectories must be such that each point $(\vec x, t)$ in the support of the wavefunction is associated with a single particle trajectory. This is true for pilot-wave trajectories, as the guidance equation is based on configuration space instead of phase space. Third, the effect of $e_I$ is to scale (amplify or suppress) the magnitude of the partial wavepackets in the probability density in a trajectory-dependent manner. Suppose the particle crosses slit A (B) to reach the coordinate $(\vec x, t)$, then the contribution of the empty wavepacket $\psi_B(\vec x, t)$ ($\psi_A(\vec x, t)$) gets scaled by the $e_I$ term according to (\ref{wow}). Fourth, if $e_I = 0$, then only the effect of the charge $e$ on the phase of the empty wavepacket in (\ref{wow}) remains. As only relative phases can be observed, the probabilities become insensitive to the which-way information and the usual AB predictions are recovered. 

\subsection{Experimental Proposal}\label{prapojal}
Let us discuss if an experiment can be set up to test equation (\ref{wow}).
As $e_I/e \approx 10^{-21}$ for electrons, the effect of the scale factor $e^{\pm \frac{e_I}{\hbar c}\Phi}$ ($\Phi \equiv \oint_\mathcal{C} \vec A(\vec x') \cdot d \vec x' $) in (\ref{wow}) is likely to be observable only in extreme regimes. We know that the flux needed to experimentally observe a $2\pi$ phase difference in the standard AB effect is given by $\Phi_q =hc/e \approx 4.14 \times 10^{-7} \text{Gcm}^2$. In our case, $\Phi_q$ produces a very large scale factor $e^{2\pi} \approx 535$. To experimentally observe a 10 $\%$ change in amplitude scale, which corresponds to a scale factor of $1.1$, we need 
\begin{align}
   & \frac{m \sqrt G \Phi}{\hbar c} = \ln (1.1) \nonumber \\
    \Rightarrow& m\Phi \approx 1.16 \times 10^{-14} \text{g esu} \label{flux}
\end{align}
where we have used equation (\ref{mast}).\\ 

\begin{figure}
    \centering
    \includegraphics[scale=0.5]{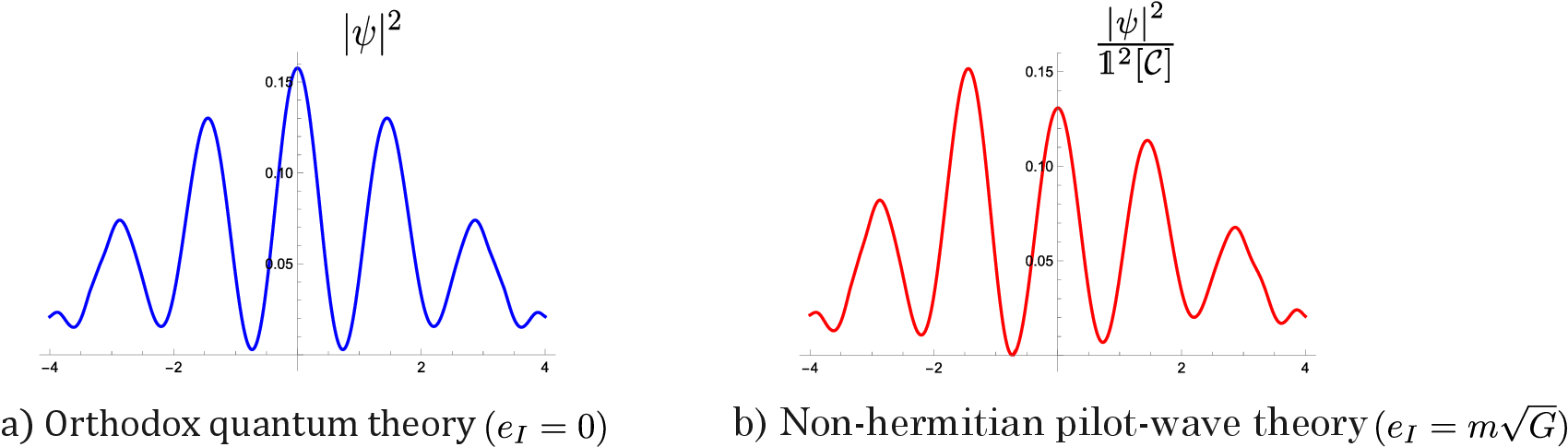}
    \caption{Schematic comparison of probability density in a) orthodox quantum theory and b) non-Hermitian PWT in Aharonov-Bohm double-slit experiment for a neutral, heavy molecule. Orthodox quantum theory predicts that the probability density $|\psi|^2$ would be identical to the standard double-slit experiment as the particle is electrically neutral. In non-Hermitian PWT, the particle couples geometrically to the gauge field $A^\mu$ via its mass, which amplifies (suppresses) the probability density $|\psi|^2/\mathds 1^2[\mathcal C]$ in the region $x \leq -0.7$ ($x > -0.7$) via the scale factor $e^{\frac{m\sqrt G}{\hbar c}\oint_{\mathcal{C}_L} \vec A(\vec x') \cdot d \vec x'}$ ($e^{-\frac{m\sqrt G}{\hbar c}\oint_{\mathcal{C}_L} \vec A(\vec x') \cdot d \vec x'}$) (see equation \ref{woho}). The point $x = -0.7$ is chosen for illustration as the separatrix, such that points to its left (right) have particle trajectories incoming from the left (right) slit. The initial wavefunction passing through the slits is taken to be $\psi(x, y, 0) = (8/\pi)(e^{-8(x-1.5)^2} + e^{-8(x+1.5)^2})e^{-8y^2}e^{i5y}$, and the probability densities are shown at $(t = 0.7, y=3.5)$. We have taken $m = G = \hbar = c = 1$ and $\oint_{\mathcal{C}_L} \vec A(\vec x') \cdot d \vec x' = \pi/4$ for illustration.}
    \label{direkt}
\end{figure}

Equation (\ref{flux}) suggests performing the following experiment. Consider a double-slit AB experiment with an electrically-neutral large molecule having mass $\sim 10^{-19}$ g, such as used in matter-wave interferometry experiments \cite{f1, f2, f3}. Suppose the magnetic flux is very high $\sim 10^5 \text{ esu}$, corresponding to a magnetic field $\sim 10^6 \text{ G}$ ($100 \text{ T}$) and solenoidal area $\sim 0.1 \text{ cm}^2$, so that the condition (\ref{flux}) is satisfied. In this regime, orthodox quantum theory and non-Hermitian PWT predict two different interference patterns. Orthodox quantum theory predicts the standard double-slit interference pattern as the molecule, being neutral, does not couple to the gauge field. Non-Hermitian PWT predicts that the particle geometrically couples to the gauge field via its mass, which modifies the probability density to
\begin{align}
   \frac{|\psi'_{AB}(\vec x, t)|^2}{\mathds 1^2[\mathcal C]} = \left \{ \begin{array}{cc}
        |\psi_A(\vec x, t) + \psi_B(\vec x, t)e^{-\frac{m\sqrt G}{\hbar c}\oint_{\mathcal{C}_L} \vec A(\vec x') \cdot d \vec x'}|^2/2  &, \text{ if the particle trajectory crosses slit A} \\
        \textbf{ } & \textbf{ } \\
        |\psi_A(\vec x, t)e^{\frac{m\sqrt G}{\hbar c}\oint_{\mathcal{C}_L} \vec A(\vec x') \cdot d \vec x'} + \psi_B(\vec x, t)|^2/2  &, \text{ if the particle trajectory crosses slit B}
    \end{array} \right . \label{woho}
\end{align}
The difference in the experimental predictions is schematically shown in figure \ref{direkt}. Note that, in a double-slit experiment, there exists a separatrix on the screen such that points to its left (right) have particle trajectories incoming from the left (right) slit \cite{1979wow, kaye, hollandbook}. The separatrix can be used to simplify the evaluation of the probability density (\ref{woho}) for computational purposes. For zero flux ($\Phi = 0$), the separatrix lies symmetrically between the two slits. For non-zero flux and $e=0, e_I \neq 0$, as in equation (\ref{woho}), the separatrix shifts towards the slit with smaller partial wavepacket magnitude. 

\subsection{Comparison with other quantum formulations} \label{compare}
The pilot-wave prediction is dependent on the which-way information due to the trajectory-dependent factor $\mathds 1[\mathcal C]$ in (\ref{woho}). Mathematically, the probability density at time $t$ is a \textit{spacetime functional} of the quantum state. This means that any quantum formulation where the probabilities at time $t$ are encoded as function of the quantum state at time $t$ cannot reproduce the prediction (\ref{woho}). Therefore, the proposed experiment in section \ref{prapojal} can potentially allow us to distinguish between non-Hermitian PWT and other quantum formulations. Let us discuss some of the formulations below.\\

First, consider quantum formulations where Hermiticity is considered a basic postulate, such as orthodox quantum theory. Such formulations predict that even at extremely high flux regimes, there would be no non-integrable scale effects as $e_I$ is necessarily zero. Clearly, then, if the prediction (\ref{woho}) is observed (not observed), then the Hermitian quantum formulation (non-Hermitian PWT) will be experimentally ruled out. \\

Second, consider non-Hermitian extensions of quantum formulations. There are, in principle, multiple possible ways to generalize a quantum formulation to allow non-Hermiticity and accommodate a non-zero $e_I$. Let us consider, for example, non-Hermitian extensions of formulations where the pilot-wave trajectories effectively appear as average trajectories, devoid of ontological status. Such formulations have been proposed, for example, based on weak values \cite{koki, fools} and stochastic hydrodynamical models \cite{neelu}. In these extensions, a spacetime point in the support of the wavefunction is not associated with a single particle trajectory at the ontological level. Instead, the formulation assigns a probability $p_A$ ($p_B$) to the averaged particle trajectory crossing slit A (B), where $p_A + p_B = 1$. If the formulation assigns the probability density for each case as given in (\ref{woho}), then the average probability density would then be given by
\begin{align}
    \rho(\vec x, t) = p_A \frac{|\psi_A(\vec x, t) + \psi_B(\vec x, t)e^{-\frac{m\sqrt G}{\hbar c}\oint_\mathcal{C} \vec A(\vec x') \cdot d \vec x'}|^2}{2} + p_B  \frac{|\psi_A(\vec x, t)e^{\frac{m\sqrt G}{\hbar c}\oint_\mathcal{C} \vec A(\vec x') \cdot d \vec x'} + \psi_B(\vec x, t)|^2}{2} \label{counter}
\end{align}
at each point $(\vec x, t)$. Clearly, such non-Hermitian extensions cannot reproduce the prediction (\ref{woho}), which differs from (\ref{counter}). \\

Consider non-Hermitian extensions based on post-selection \cite{loyd1, loyd2}. These renormalize the quantum state by dividing it with the changed norm. However, such a renormalization factor is global, whereas the factor $\mathds 1 [\mathcal C]$ is local (varies with the spacetime point). Therefore, they cannot reproduce (\ref{woho}). \\

Let us, lastly, consider hypothetical non-Hermitian extensions that reproduce the pilot-wave prediction (\ref{woho}) without seemingly using pilot-wave trajectories. Although no such formulation is currently available, it is possible to imagine one in the future. Such formulations would uniquely associate each spacetime point in the support of the wavefunction with one of the expressions in (\ref{woho}), but would not describe the case structure in (\ref{woho}) in terms of the which-way information. The underlying mathematics in the hypothetical formulation to describe the case structure, however, must be the same as or equivalent to the which-way information in PWT for consistency. This implies that the difference would be merely semantic, since the which-way information in PWT would correspond to whatever terminology is used to describe the cases in (\ref{woho}) in the hypothetical formulation. \\

Our list of hypothetical non-Hermitian extensions is not exhaustive. It is always possible to construct theories that fit experimental data. However, the pilot-wave prediction cannot be reproduced by any other quantum formulation that does not use the same trajectories (or their mathematical equivalent) to generate its predictions, or those formulations that do not have trajectories (or their mathematical equivalent). Unlike in the Hermitian case, particle trajectories cannot be considered ``hidden variables'' with no connection to experimental predictions. 

\section{Assessment of the Weyl-Einstein debate} \label{wade}
Einstein gave an early criticism of Weyl's gauge theory in 1918 \cite{einstein18, gawn, vizgin, ryckmanbook}. The criticism, further elaborated by Pauli \cite{pauli58}, was considered devastating for the physical viability of Weyl's theory. Interestingly, Einstein's criticism is conceptually related to aspects of quantum theory, which was as yet undeveloped in 1918. Later, as the orthodox formulation of quantum theory took shape over the 1920s, its Hermiticity postulate compelled a reinterpretation of Weyl's gauge symmetry idea as local phase invariance in the theory \cite{fuck, london27, gawn}. It has, therefore, remained impossible to clearly analyse Einstein's criticism, related to local scale invariance, at the quantum level. \\

Einstein's criticism may be summarized by the claim that spectral frequencies in a local scale invariant theory are history dependent. The observed constancy of spectral frequencies then empirically contradicts such a theory. Weyl's response to the criticism evolved over time, but in general he argued that it is necessary to first formulate local scale invariant equations of motion, missing in Einstein's criticism, to extract experimental predictions \cite{weyl18, weyl22, vizgin, ryckmanbook}. Below we show that Einstein's criticism is incorrect, by analysing the absorption spectrum of a local scale invariant quantum harmonic oscillator interacting with light using non-Hermitian PWT. \\

\subsection{Local scale invariant harmonic oscillator with complex frequency} \label{complex}
Consider a non-relativistic electron in a 3D harmonic-oscillator potential that is effectively configured by coupling the electron with the scalar potential $\phi(\vec x) = (\lambda^2_x x^2 + \lambda^2_y y^2 + \lambda^2_z z^2)/2$, where $\lambda_x, \lambda_y, \lambda_z$ are coupling constants of appropriate dimensions. The vector potential is set $\vec A(\vec x) = 0$. The time-independent Schrödinger equation is 
\begin{align}
     -\frac{\hbar^2}{2m}\vec \nabla^2\psi(\vec{x}) + e_C \phi(\vec x)\psi(\vec{x}) = E \psi(\vec{x}) \nonumber \\
     \Rightarrow -\frac{\hbar^2}{2m}\vec \nabla^2 + \frac{1}{2}m \vec \omega^2 \psi(\vec{x}) = E \psi(\vec{x})
\end{align}
where the oscillator has a complex frequency 
\begin{align}
    \vec \omega = (\sqrt{e_C\lambda_x^2/m}, \sqrt{e_C\lambda_y^2/m}, \sqrt{e_C\lambda_z^2/m}) \label{freq}
\end{align}
It is straightforward to show (see appendix \ref{jaiho}) that the usual method of solving for the harmonic oscillator eigenstates and eigenvalues can be directly applied to this case. The energy eigenvalues are given by
\begin{align}
    E_{\vec n} = \hbar \sum_j \omega_j (n_j+1/2) \label{energy}
\end{align}
where $j \in \{x, y, z\}$, $\vec n \equiv (n_x, n_y, n_z)$ and $n_j \in \{0, 1, 2...\}$ labels the $n_j^{th}$ harmonic-oscillator energy level along the direction $j$. Clearly, the energies are complex like $\vec \omega$. Splitting $\vec \omega$ into its real and imaginary components $\vec \omega = \vec \omega^R + \vec \omega^I$, the energies can be rewritten as 
\begin{align}
        E_{\vec n} = E_{\vec n}^R + i E_{\vec n}^I  
\end{align}
where $E_{\vec n}^R \equiv \hbar \sum_j \omega_j^R (n_j+1/2)$ and $E_{\vec n}^I \equiv \hbar \sum_j \omega_j^I (n_j+1/2)$.
As $e_I/e \sim 10^{-21}$ for electrons (see section \ref{value}), $\omega^I_j \ll \omega^R_j $ $\forall j$.  \\

The corresponding energy eigenstates $|\vec n \rangle$ are the usual harmonic-oscillator eigenstates with complex $\vec \omega$, given in the position representation by $ \langle \vec x |{\vec n}\rangle = \phi_{\vec n} (\vec x) = \Pi_j \phi_{n_j}(j)$, where
\begin{align}
   \phi_{n_j}(j) = \mathcal N_{n_j}e^{-\frac{m\omega_j j^2}{2\hbar}}H_{n_j}(\sqrt{\frac{m\omega_j}{\hbar}}\textbf{ }j) \label{tis}
\end{align}
Here $H_{n_j}(j)$ is the $n_j^{th}$ Hermite function in coordinate $j \in \{x, y, z\}$ and $\mathcal{N}_{n_j}$ is a normalization constant.

\subsection{Spectroscopic energy measurement}\label{specter}
Suppose the electron starts interacting with a monochromatic electromagnetic wave of frequency $\omega$, propagating along $\hat{k}$ with linear polarization along $\hat \epsilon$. The vector potential desribing the wave is given by
\begin{align}
    \vec A(\vec x, t) &= \hat \epsilon A_0 \cos \big ( \frac{\omega }{c}\hat k \cdot \vec x - \omega t\big ) \textbf{ } \label{plane}
\end{align}
where $\hat \epsilon \cdot \hat k = 0$. The Schrödinger equation describing the electron is given by \cite{sen26a}
\begin{align}
     \bigg(-\frac{\hbar^2}{2m}\vec \nabla^2 +e_C \phi(\vec x)+\frac{i\hbar e_C}{mc} A_0\cos \big ( \frac{\omega }{c}\hat n \cdot \vec x - \omega t\big )\textbf{ } \hat \epsilon\cdot \vec \nabla \bigg)\psi(\vec{x},t) = i \hbar \frac{\partial \psi(\vec{x},t)}{\partial t}
\end{align}
where we have assumed that $A_0$ is small enough so that $A_0^2$ terms can be ignored in the Hamiltonian. The total Hamiltonian can be split into $\hat H_0 + \hat H'(t)$ where $\hat H_0 \equiv -(\hbar^2/2m)\vec \nabla^2 + e_C \phi(\vec x)$ and $\hat H'(t) \equiv (i\hbar e_C/mc) A_0 \cos \big ( \frac{\omega }{c}\hat n \cdot \vec x - \omega t\big )\textbf{ }\hat \epsilon\cdot \vec \nabla$. We treat $\hat H_0$ as the unperturbed Hamiltonian, and treat $\hat H'(t)$ as a time-dependent perturbation that can cause transitions between the energy levels $E_{\vec n}$ of $\hat H_0$ given by (\ref{energy}). We assume for convenience that $E_{\vec n}^R$ is non-degenerate. Note that, as $e_C$ is complex, both $\hat H$ and $\hat H'(t)$ are non-Hermitian.\\

\begin{figure}
    \centering
    \includegraphics[width=0.5\linewidth]{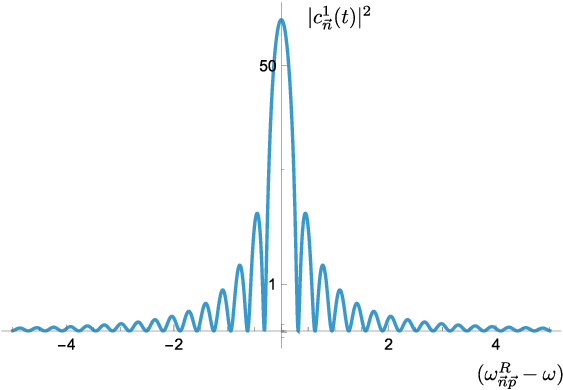}
    \caption{History-independence of spectral frequencies. The plot shows that the resonance condition $\omega^R_{\vec n\vec p} = \omega $ for maximizing $|c_{\vec n}^1(t)|^2$ is unaffected by the history of the particle. This implies that Einstein's criticism of local scale invariance that it leads to history-dependent spectral frequencies (``second-clock effect'') is incorrect in non-Hermitian PWT. We have taken $\omega^I_{\vec n \vec p}/\omega^R_{\vec n \vec p} \approx 10^{-21}$, $\overline{V}_{\vec n \vec p} = 1$ and $t = 10$ in equation \eqref{tdptm} for illustration.}
    \label{fig1}
\end{figure}

From standard results in the interaction picture, which are easily verifiable to remain valid for non-Hermitian hamiltonians, the quantum state at time $t$ is given by
\begin{align}
    |\psi(t)\rangle = \sum_{\vec n} c_{\vec n}(t) e^{(-iE_{\vec n}^R + E_{\vec n}^I)t/\hbar} |{\vec n}\rangle \label{td1}
\end{align}
where $\hat H_0 |{\vec n}\rangle = (E_{\vec n}^R + i E_{\vec n}^I)|{\vec n}\rangle$ and $|{\vec n}\rangle$ are given by (\ref{tis}). Suppose the initial state of the system (before the interaction) is $|\vec p \rangle$. The first order approximation to $|c_{\vec n}(t)|^2$ can then be shown to be (see appendix \ref{jago})
\begin{align}
    | c_{\vec n}^1(t)|^2 \approx \frac{ |e_C|^2 A_0^2}{m^2 c^2} \times \left \{ \begin{array}{cc}
      \frac{ \sin^2{(\omega^R_{\vec n\vec p}+\omega)t}+(\omega^I_{\vec n\vec p}t)^2  }{(\omega^R_{\vec n\vec p}+ \omega)^2 + (\omega^I_{\vec n\vec p})^2}|V_{\vec n \vec p}|^2 & , \text{if }\omega^R_{\vec n \vec p}+ \omega \approx 0  \\~\\
      \frac{ \sin^2{(\omega^R_{\vec n\vec p}-\omega)t}+(\omega^I_{\vec n\vec p}t)^2 }{(\omega^R_{\vec n\vec p}- \omega)^2 + (\omega^I_{\vec n\vec p})^2} |\overline{V}_{\vec n \vec p}|^2 & , \text{if } \omega^R_{\vec n \vec p}- \omega \approx 0 \label{tdptm}
    \end{array} \right .
\end{align}
where we have assumed $(\omega^I_{\vec n\vec p}t$) to be small. The second case of equation (\ref{tdptm}) is illustrated in figure \ref{fig1}.\\

Note that $| c_{\vec n}^1(t)|^2$ is \textit{not} the transition probability. Using equations \eqref{tis}, (\ref{td1}) and the conservation of the density $|\psi(\vec x, t)|^2/\mathds 1^2[\mathcal C]$, the probability of transition from $|\vec p\rangle \to |\vec n\rangle$ is given by
\begin{align}
    P_{\vec p \to \vec n}(t) &\equiv | c_{\vec n}^1(t)|^2 e^{2E_{\vec n}^I t/\hbar}\int\frac{|\phi_{\vec n}(\vec x)|^2}{\mathds 1^2[\mathcal C]} d^3 \vec x \label{teri}
\end{align}
The integral in (\ref{teri}) is history-dependent via the term $\mathds 1[\mathcal C]$. As the time scale of interaction in a spectroscopic measurement is of a very short duration compared to the electron's history, $\mathds 1[\mathcal C]$ effectively depends on only the history before the interaction. Therefore, $e^{2E_{\vec n}^I t/\hbar}\int |\phi_{\vec n}(\vec x)|^2/\mathds 1^2[\mathcal C] d^3 \vec x$ is an $\omega$-independent \textit{scale} multiplying $| c_{\vec n}^1(t)|^2$. Clearly, as figure \ref{fig1} illustrates, the resonance frequencies are then determined exclusively from $| c_{\vec n}^1(t)|^2$, which is maximized when $\omega = \pm \omega^R_{\vec n \vec p}$ according to (\ref{tdptm}). Thus, we get the same resonance frequencies as in the Hermitian case, which implies that spectral frequencies are history-independent. Therefore, there is no second-clock effect in our theory despite the local scale invariance, as anticipated by Weyl.\\

\subsection{History-dependence of spectral intensities}\label{intense}
Equations \eqref{tdptm}, (\ref{teri}) imply that, although spectral frequencies are history-independent, transition probabilities are affected by the system history encoded in $\mathds 1[\mathcal C]$. This implies that the intensities of the spectral lines are history-dependent. The history-dependence would be very small in magnitude but in principle measurable. As $e_I \propto m$ (see section \ref{value}), this prediction can be realistically tested experimentally for very heavy quantum particles.

\subsection{Contribution to spectral linewidth}\label{physical}
In the long-time limit, $| c_{\vec n}^1(t)|^2$ can be further approximated as (see appendix \ref{jago})
\begin{align}
    | c_{\vec n}^1(t)|^2 \approx \frac{ |e_C|^2 A_0^2}{2  m^2 c^2} \times \left \{ \begin{array}{cc}
      \frac{ \cosh{2\omega^I_{\vec n\vec p}t}  }{(\omega^R_{\vec n\vec p}+ \omega)^2 + (\omega^I_{\vec n\vec p})^2}|V_{\vec n \vec p}|^2 & , \text{if }\omega^R_{\vec n \vec p}+ \omega \approx 0  \\~\\
      \frac{\cosh{2\omega^I_{\vec n\vec p}t} }{(\omega^R_{\vec n\vec p}- \omega)^2 + (\omega^I_{\vec n\vec p})^2} |\overline{V}_{\vec n \vec p}|^2 & , \text{if } \omega^R_{\vec n \vec p}- \omega \approx 0
    \end{array} \right .
\end{align}
which implies that
\begin{align}
    | c_{\vec n}^1(t)|^2 \propto \frac{1 }{(\omega^R_{\vec n\vec p}- \omega)^2 + (\omega^I_{\vec n\vec p})^2}
\end{align}
which is a Cauchy-Lorentz density. Clearly, $2\omega^I_{\vec n\vec p}$ is the width (full width at half maximum) of the density, corresponding to the linewidth in the absorption spectrum. Therefore, the difference in the imaginary frequencies contributes to spectral linewidths. This contribution would be small but can, in principle, be experimentally measured. Note that, unlike linewidths, we cannot directly infer the effect of imaginary frequencies on decay rates as the state norm depends on $\mathds 1[\mathcal C]$ \cite{sen26a}.

\section{Discussion} \label{disc}
We have obtained the relation $e_I = m \sqrt G$ for the imaginary component of the gauge coupling parameter by generalizing the fine-structure constant $\alpha$, which governs non-integrable phase effects, to $\alpha_S$, which governs non-integrable scale effects. The dependence of $e_I$ on gravitational quantities implies, in the context of Weyl's unified field theory, that particles geometrically couple to the electromagnetic gauge field, which affects the scaling geometry of the spacetime manifold. This geometric coupling is distinct from the charge-dependent dynamical coupling. For the electron, we found $e_I$ to be $\sim 10^{-21}$ times the electronic charge, which helps explain why non-integrable scale effects can be ignored in usual quantum experiments. \\

Non-Hermitian PWT makes an unique experimentally-testable prediction in an Aharonov-Bohm double-slit experiment. It predicts that the probability density at a spacetime point explicitly depends on the particle trajectory passing through it (see equation (\ref{wow})). This is because the contribution of the empty wavepacket to the probability density is modified by a gauge-invariant scale factor that depends on which slit the particle trajectory crosses. We have argued that the prediction cannot be reproduced by other quantum formulations that do not use pilot-wave trajectories, or their mathematical equivalents, to determine probabilities (see section \ref{compare}). The prediction can be tested for a heavy, neutral molecule with mass $m \sim 10^{-19}$ g and a solenoidal flux $10^5$ esu. However, currently available experimental technology severely limits the ability to perform such an experiment. Therefore, future work aimed at both improving the technology and theoretically optimizing the experimental proposal is required.\\

We have used non-Hermitian PWT to show that Einstein's criticism of Weyl's unified field theory, that local scale invariance leads to history-dependent spectral frequencies (second-clock effect) \cite{einstein18}, is incorrect. Einstein's argument, made nearly a decade before quantum theory had clearly emerged, implicitly assumes that the details of quantum formulations are unnecessary when discussing spectral frequencies. Weyl contested this, writing, ``the mathematical ideal of vector transfer...has nothing to do with the real situation regarding the movement of a clock, \textit{which is determined by the equations of motion}'' [emphasis added] \cite{weyl18}. We determined the behaviour of atomic clocks by analysing spectral frequencies using non-Hermitian PWT. Our analysis shows that Weyl's response was correct, and suggests that the abandonment of his theory was premature. Lastly, our work suggests that the historical role of gauge invariance in shaping the field of quantum foundations has been overlooked. Had Weyl's gauge idea been implemented at the quantum level using PWT in the 1920s, instead of orthodox quantum theory \cite{fuck, london27, vizgin, gawn}, the role of particle trajectories in formulating the probability rule might have been clearer from the outset, and particle configurations might not have acquired the label ``hidden variables''.\\

Our analysis of the Weyl-Einstein debate yielded further insights and experimental predictions. First, the intensities of spectral lines, unlike spectral frequencies, turn out to be history dependent. Second, the imaginary components of energy eigenvalues contribute to spectral linewidths. Both the history dependence of spectral intensities and the contribution to linewidths constitute predictions that can, in principle, be subject to experimental tests. \\

After the bulk of our work was completed, we came to know of several previous works \cite{suntomato84, suntomato85, suntomato14, shajai98, shajai4} that discuss local scale invariance connected to a pilot-wave context from an excellent review article on local scale invariance \cite{scholz18}. It would be interesting to study the relation, if any, between our theory and these works. We briefly note here a key distinction: our theory predicts a trajectory-dependent probability density, which renders it predictively inequivalent to orthodox quantum theory. 

\section*{Author Contributions}
Indrajit Sen: conceptualization (lead); methodology (lead); formal analysis (lead); supervision (equal); validation (equal); writing - original draft (lead); writing - review \& editing (equal). Matt Leifer: conceptualization (supporting); methodology (supporting); supervision (equal); validation (equal); writing - review \& editing (equal).

\acknowledgments
IS is thankful to Adbhut Gupta and Herman Batelaan for helpful discussions. ML was supported by Grant 63209 from the John Templeton Foundation. The opinions expressed in this publication are those of the authors and do not necessarily reflect the views of the John Templeton Foundation.

\bibliographystyle{bhak}
\bibliography{bib}

\appendix

\section{Weyl-Einstein debate} \label{wado}
\subsection{Local scale invariant harmonic oscillator with complex frequency} \label{jaiho}
The time-independent Schrödinger equation for the harmonic oscillator with a complex $\omega$ can be written as
\begin{align}
    -\frac{d^2\psi}{dy^2} + y^2\psi = K\psi \label{ho}
\end{align}
where $y\equiv \sqrt{m\omega/\hbar}x$ and $K \equiv 2E/\hbar\omega$. Note that $y$, $K$ are complex. Using the ansatz $e^{-y^2/2}h^K(y)$ in equation (\ref{ho}), we get the Hermite differential equation
\begin{align}
    \frac{d^2 h^K}{dy^2} -2y\frac{dh^K}{dy} + (K-1)h^K = 0 \label{her}
\end{align}
The general solution to (\ref{her}) can be written as
\begin{align}
     h^K(y) &= \sum_{n=0}^{\infty} a_n y^n \label{solb}
 \end{align}    
 where $a_0$ and $a_1$ are two arbitrary complex constants and the recurrence relation between $a_n$'s can be obtained to be $a_{n+2} = (2n+1-K)a_n/(n+1)(n+2)$. For the solution to normalizable, the power series must terminate. Therefore, we must have 
 \begin{align}
     2n+1 = K
 \end{align}
which yields the complex energies
\begin{align}
    E_n = (n+\frac{1}{2})\hbar \omega 
\end{align}
where $n \in \{0, 1, 2, 3...\}$.
\subsection{Non-Hermitian time-dependent perturbation theory} \label{jago}
The time-dependent coefficients $c_{\vec n}(t)$ in equation \eqref{td1} in the main text are given by
\begin{align}
    i\hbar \dot c_{\vec n}(t) = \sum_m H'_{\vec n\vec m}(t) e^{i\omega^R_{\vec n\vec m}t - \omega^I_{\vec n \vec m}t}c_{\vec m}(t) \label{td2}
\end{align}
Here $\omega^R_{\vec n\vec m} \equiv (E_{\vec n}^R -E_{\vec m}^R)/\hbar$, $\omega^I_{\vec n\vec m} \equiv (E_{\vec n}^I -E_{\vec m}^I)/\hbar > 0$ and  
\begin{align}
    H'_{\vec n\vec m}(t) \equiv \langle \vec n|\hat H'(t) |\vec m \rangle = \int \overline{\phi_{\vec n}}(\vec x) \hat{H}'(t) \phi_{\vec m}(\vec x) \textbf{ }d^3 \vec x \label{tdef}
\end{align}
where $\langle \vec x |{\vec n}\rangle = \phi_{\vec n}(\vec x)$ is defined by \eqref{tis} in the main text. The coefficients $c_{\vec n}(t)$ can be approximated by time-dependent perturbation theory using Dyson series, which is easily verifiable to be valid\footnote{The Dyson series is also valid for non-normalizable quantum states, see \cite{sen22}.} for non-Hermitian $\hat H'$. Suppose the system is in the eigenstate $\phi_{\vec p}(\vec x)$ at time $-t$ and the perturbation acts during the time interval $(-t, +t)$. The first order approximation to $c_{\vec n}(t)$ is then given by
\begin{align}
    c_{\vec n}^1(t)  = \frac{-i}{\hbar}\int_{-t}^{t} e^{i\omega^R_{\vec n\vec p}t' - \omega^I_{\vec n \vec p}t'}\textbf{ } H'_{\vec n\vec p}(t') \textbf{ } dt' \label{td3}
\end{align}
Substituting $\hat H'(t) =(i\hbar e_C/mc)A_0(\omega) \cos \big ( \frac{\omega }{c}\hat k \cdot \vec x - \omega t\big )\textbf{ } \hat \epsilon\cdot \vec \nabla$ in equation (\ref{td3}), we get
\begin{align}
    c_{\vec n}^1(t) = \frac{e_C A_0}{2mc} \bigg (\frac{e^{i(\omega^R_{\vec n\vec p}+\omega)t-\omega^I_{\vec n\vec p}t}-e^{-i(\omega^R_{\vec n\vec p}+\omega)t+\omega^I_{\vec n\vec p}t}}{i(\omega^R_{\vec n\vec p}+ \omega) - \omega^I_{\vec n\vec p}} V_{\vec n\vec p} + \frac{e^{i(\omega^R_{\vec n\vec p}-\omega)t-\omega^I_{\vec n\vec p}t}-e^{-i(\omega^R_{\vec n\vec p}-\omega)t+\omega^I_{\vec n\vec p}t}}{i(\omega^R_{\vec n\vec p}- \omega) - \omega^I_{\vec n\vec p}} \overline{V}_{\vec n \vec p}  \bigg) \label{tdpt}
\end{align}
where $V_{\vec n \vec p}$, $\overline{V}_{\vec n \vec p}$ are defined by equation (\ref{tdef}) and $V \equiv e^{i\frac{\omega}{c}\hat k \cdot \vec x}$. Note that $\overline{V}_{\vec n \vec p} \neq \overline{V_{\vec n \vec p}}$. From (\ref{freq}) in the main text, given that $e_I/e \sim 10^{-21}$ for electrons, it is straightforward to show that $\omega^I_{\vec n \vec p}/\omega^R_{\vec n \vec p} \sim 10^{-21}$. The magnitude of the first (second) term in brackets in (\ref{tdpt}) is significantly larger than the other whenever $\omega^R_{\vec n\vec p}+\omega \approx 0$ ($\omega^R_{\vec n\vec p}-\omega \approx 0$) as $\omega^I_{\vec n\vec p}$ is very small. Therefore, equation (\ref{tdpt}) implies that 
\begin{align}
    | c_{\vec n}^1(t)|^2 \approx \frac{ |e_C|^2 A_0^2}{2  m^2 c^2} \times \left \{ \begin{array}{cc}
      \frac{ \cosh{2\omega^I_{\vec n\vec p}t} - \cos{2(\omega^R_{\vec n\vec p}+\omega)t} }{(\omega^R_{\vec n\vec p}+ \omega)^2 + (\omega^I_{\vec n\vec p})^2}|V_{\vec n \vec p}|^2 & , \text{if }\omega^R_{\vec n \vec p}+ \omega \approx 0  \\~\\
      \frac{\cosh{2\omega^I_{\vec n\vec p}t} - \cos{2(\omega^R_{\vec n\vec p}-\omega)t}}{(\omega^R_{\vec n\vec p}- \omega)^2 + (\omega^I_{\vec n\vec p})^2} |\overline{V}_{\vec n \vec p}|^2 & , \text{if } \omega^R_{\vec n \vec p}- \omega \approx 0 \label{tdpt0}
    \end{array} \right .
\end{align}
We can further approximate (\ref{tdpt0}) at small $t$ as
\begin{align}
    | c_{\vec n}^1(t)|^2 \approx \frac{ |e_C|^2 A_0^2}{m^2 c^2} \times \left \{ \begin{array}{cc}
      \frac{ \sin^2{(\omega^R_{\vec n\vec p}+\omega)t}+(\omega^I_{\vec n\vec p}t)^2  }{(\omega^R_{\vec n\vec p}+ \omega)^2 + (\omega^I_{\vec n\vec p})^2}|V_{\vec n \vec p}|^2 & , \text{if }\omega^R_{\vec n \vec p}+ \omega \approx 0  \\~\\
      \frac{ \sin^2{(\omega^R_{\vec n\vec p}-\omega)t}+(\omega^I_{\vec n\vec p}t)^2 }{(\omega^R_{\vec n\vec p}- \omega)^2 + (\omega^I_{\vec n\vec p})^2} |\overline{V}_{\vec n \vec p}|^2 & , \text{if } \omega^R_{\vec n \vec p}- \omega \approx 0 \label{tdpt1}
    \end{array} \right .
\end{align}
where we have used $\cosh{2\omega^I_{\vec n\vec p}t} \approx 1 + 2(\omega^I_{\vec n\vec p}t)^2$, assuming $t$ is such that $\omega^I_{\vec n\vec p} t$ is very small. In the long-time limit, equation \eqref{tdpt0} can be approximated as 
\begin{align}
    | c_{\vec n}^1(t)|^2 \approx \frac{ |e_C|^2 A_0^2}{2  m^2 c^2} \times \left \{ \begin{array}{cc}
      \frac{ \cosh{2\omega^I_{\vec n\vec p}t}  }{(\omega^R_{\vec n\vec p}+ \omega)^2 + (\omega^I_{\vec n\vec p})^2}|V_{\vec n \vec p}|^2 & , \text{if }\omega^R_{\vec n \vec p}+ \omega \approx 0  \\~\\
      \frac{\cosh{2\omega^I_{\vec n\vec p}t} }{(\omega^R_{\vec n\vec p}- \omega)^2 + (\omega^I_{\vec n\vec p})^2} |\overline{V}_{\vec n \vec p}|^2 & , \text{if } \omega^R_{\vec n \vec p}- \omega \approx 0
    \end{array} \right .
\end{align}

\end{document}